\documentclass{IEEEtran4PSCC}
\ifCLASSINFOpdf
   \usepackage[pdftex]{graphicx}
   \usepackage{tikz}
\else
   \usepackage[dvips]{graphicx}
\fi
\usepackage[cmex10]{amsmath}

\usepackage{cite}
\usepackage{amsfonts}
\usepackage{siunitx}

\ifCLASSOPTIONcompsoc
 \usepackage[caption=false,font=normalsize,labelfont=sf,textfont=sf]{subfig}
\else
 \usepackage[caption=false,font=footnotesize]{subfig}
\fi
\usepackage{stfloats}

\makeatletter
\let\old@ps@headings\ps@headings
\let\old@ps@IEEEtitlepagestyle\ps@IEEEtitlepagestyle
\def\psccfooter#1{%
    \def\ps@headings{%
        \old@ps@headings%
        \def\@oddfoot{\strut\hfill#1\hfill\strut}%
        \def\@evenfoot{\strut\hfill#1\hfill\strut}%
    }%
    \def\ps@IEEEtitlepagestyle{%
        \old@ps@IEEEtitlepagestyle%
        \def\@oddfoot{\strut\hfill#1\hfill\strut}%
        \def\@evenfoot{\strut\hfill#1\hfill\strut}%
    }%
    \ps@headings%
}
\makeatother

\psccfooter{%
        \parbox{\textwidth}{\hrulefill \\ \small{24th Power Systems Computation Conference} \hfill \begin{minipage}{0.2\textwidth}\centering \vspace*{4pt} \includegraphics[scale=0.06]{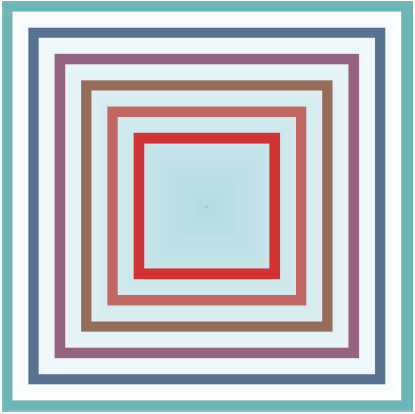}\\\small{PSCC 2026} \end{minipage} \hfill \small{Limassol, Cyprus --- June 8-12, 2026}}%
}

\begin{document}
%
\title{Risk-Aware Multi-Market Scheduling of Virtual Power Plants with Dynamic Network Tariffs}

\author{
\IEEEauthorblockN{Lorenzo Zapparoli, Paul Fäth, Blazhe Gjorgiev, Giovanni Sansavini}
\IEEEauthorblockA{Institute of Energy and Process Engineering \\
ETH Zurich\\
Zurich, Switzerland\\
\{lzapparoli, faethp, gblazhe, sansavig\}@ethz.ch}
}


\maketitle

\begin{abstract}
As the penetration of distributed energy resources (DERs) increases, harnessing their flexibility becomes critical for power system operations. Virtual power plants (VPPs) offer a promising solution. However, most existing scheduling tools rely on simplified DER or grid models and largely overlook local flexibility procurement mechanisms such as dynamic network tariffs.
This paper proposes a two-stage stochastic optimization framework for VPP multi-market scheduling that integrates detailed device-level constraints, network limitations, and operational and market uncertainties. Conditional value-at-risk is incorporated to represent risk preferences, and Benders decomposition ensures tractability with extensive scenario sets. The model jointly optimizes bidding across energy and reserve markets while explicitly accounting for local flexibility procurement through dynamic network tariffs.
The results from a realistic case study show that both risk-neutral and risk-averse strategies exploit arbitrage opportunities. However, risk aversion reduces profit volatility through closer alignment with physical dispatch. Dynamic tariffs unlock local flexibility by shifting demand across the day, though strong tariff signals reduce expected profitability by up to 65\% with limited additional flexibility gains.
\end{abstract}


\thanksto{\noindent Submitted to the 24th Power Systems Computation Conference (PSCC 2026).}

\section{Introduction}
The transition toward a low-carbon energy system is reshaping how electricity is generated, consumed, and transported. The growing integration of renewable energy sources (RES)~\cite{IRENA2022} and the rapid deployment of distributed energy resources (DERs)~\cite{IEA_Report} present both opportunities and challenges for grid operators. While these technologies support decarbonization efforts, their variability and decentralization increase the demand for system flexibility~\cite{IEA_Report}. In response, system operators seek new sources of flexibility to maintain grid stability, with virtual power plants (VPPs) emerging as a promising solution~\cite{Riaz}. By aggregating diverse DERs, such as distributed generators (DG), heat pumps (HP), electric vehicles (EV), and battery energy storage systems (BESS), VPPs operate as unified entities that can interact with electricity markets similarly to conventional power plants.

Unlike conventional centralized generators, DERs are inherently uncertain, exhibiting volatility in weather-dependent generation, thermal demand, and user-driven behavior. These uncertainties affect both revenue streams and operational feasibility, particularly when physical network constraints are considered. Thus, it is critical to co-optimize market participation and DER dispatch under uncertainty. Moreover, unlocking DER flexibility requires the coordinated participation in sequential electricity markets (e.g., energy and reserve), which adds further layers of complexity for VPP operators.

Multi-stage stochastic programming has emerged as the predominant method for modeling such intertemporal market participation under uncertainty. A three-stage program for scheduling VPPs in the day-ahead, secondary reserve capacity and activation, and imbalance markets is proposed in~\cite{Kraft2023}. The paper demonstrates how the model can design bidding strategies parameterized to the portfolio manager's risk aversion. However, to maintain computational tractability, it aggregates all the DERs in a single unit, neglects network constraints, and simulates only 250 scenarios. In~\cite{VahedipourDahraie}, the authors incorporate grid constraints into a two-stage stochastic formulation and implement Benders decomposition to achieve computational tractability. However, the VPP is represented by a simplified 15-bus network with only three PV and BESS units each, aggregating 2,000 households. This level of abstraction overlooks the operational diversity and flexibility potential of individual DERs. Similarly,~\cite{Fusco2023} develops a multi-stage program for a VPP in Italy that includes complex ancillary market structures. However, the formulation does not consider reserve capacity procurement and is demonstrated on a small test case study.
Alternative formulations such as robust optimization~\cite{NEMATI2025110594, 7742043} provide hedging against worst-case uncertainty realizations but often yield overly conservative solutions. The use of distributionally robust joint chance-constraints is also proposed~\cite{KIM2024133712}; however, the model's complexity limits the applicability to small portfolios.

While most studies focus on VPPs interacting with centralized markets, there is growing interest in exploiting their flexibility at the distribution level. Distribution system operators are exploring mechanisms such as dynamic network tariffs, which send time-varying price signals to DERs to mitigate local congestion and defer grid reinforcements. The work in~\cite{Avau2021} demonstrates that dynamic tariffs can encourage load shifting in VPPs, but the proposed model excludes market bidding and uncertainty. Similarly, studies~\cite{Ghorbankhani, He} explore tariff-based flexibility, yet rely on simplified DER representations and overlook how VPPs respond to simultaneous market and tariff signals under uncertainty.

Despite significant progress in modeling VPP participation in electricity markets, gaps remain. First, many existing approaches rely on highly aggregated representations of DERs or simplified grid topologies, which overlook the spatial, technical, and operational diversity of residential-scale VPPs. This abstraction limits the accuracy of dispatch strategies and the evaluation of grid-supportive behavior. Second, while regulatory instruments such as dynamic network tariffs are increasingly discussed as tools to unlock local flexibility, the joint impact of dynamic tariffs and market prices on VPP dispatch and profitability remains insufficiently explored.

Therefore, we aim to answer the following research questions: 
(i) How can a DER-based VPP optimally participate in sequential energy and reserve markets under price and operational uncertainty, while accounting for dynamic network tariffs? (ii) To what extent do dynamic network tariffs affect the profitability and flexibility of VPPs when considered in conjunction with market operations?

To address these questions, this paper develops a highly resolved, two-stage stochastic optimization model for the multi-market scheduling of DER-based VPPs. The model allows for the VPP participation in the energy and ancillary services (reserve capacity and reserve activation) markets. The proposed framework captures technical constraints and uncertainties at the individual device level and integrates power flow equations to account for network limitations. Financial and operational uncertainties are modeled through an extensive scenario set. Conditional value-at-risk (CVaR) is used to incorporate risk preferences into the market bidding strategy. We apply Benders decomposition to ensure tractability despite the model’s high dimensionality. This work advances the state of the art (i) by explicitly incorporating dynamic network tariffs into a risk-aware VPP scheduling model and (ii) by assessing their impact on VPP profitability and flexibility provision. The decomposition-based solution approach enables the analysis of a realistic VPP case study with detailed DER models and large scenario sets, going beyond the simplified settings commonly adopted in the literature and providing insights for market operators and policymakers.

\section{Method}
This section presents the mathematical formulation of the VPP multi-market scheduling optimization model. On the day before the delivery, the VPP operator places bids in the reserve capacity market and in the day-ahead market. Then, during the delivery day, the operator dispatches the DERs, bids in the reserve activation market, and realizes imbalances. The intraday market is not considered in this study. 

We model this decision process as a two-stage stochastic program. The operator decides in the first stage the quantities to bid in the day-ahead and the reserve capacity markets. When deciding these quantities, the VPP operator accounts for the uncertainty in the next-day forecasts of prices and DER operations. In the second stage, which represents the delivery day, the operator decides how to operate the flexible DERs, how to bid in the reserve activation market, and which imbalance position to take. Even if the reserve capacity and day-ahead markets are cleared sequentially, we model them within the same stage. This assumption is justified by the small time separation, which implies that the operator has the same information in both bidding stages~\cite{Kraft2023}.

We assume the VPP is a price taker in all markets, facing exogenous prices for energy, reserve capacity, and reserve activation. In the secondary reserve activation market, we do not model activation volume uncertainty explicitly; instead, we assume that a non-zero activation price indicates that activation occurred in the corresponding time step, allowing the VPP to define the offered volume based on price scenarios. This approach is consistent with the European market design~\cite{PICASSO}. Under these assumptions, the VPP’s decision reduces to selecting the volumes to offer in the markets, based on price scenarios.

Let $\boldsymbol{x} \in \mathbb{R}^{N^x}$ be the vector of first-stage decisions, representing energy and reserve capacity offers. The random vector $\boldsymbol{\mathrm{y}} \in \mathbb{R}^{N^y}$ comprises uncertain parameters such as market clearing prices (day-ahead, reserve, and imbalance), renewable energy availability, and load forecasts. After uncertainty is realized, the second-stage decisions $\boldsymbol{z} \in \mathbb{R}^{N^z}$ are taken and include reserve activations, imbalance quantities, and DER dispatch variables.

Based on this decision structure, the VPP operator seeks to determine first-stage bidding decisions that hedge against uncertain market and operational outcomes while accounting for risk preferences. This is achieved by minimizing a risk functional $\mathcal{R}[\cdot]$ applied to the total operating cost function $f(\boldsymbol{x}, \boldsymbol{\mathrm{y}})$, which depends on both the bidding decisions and the realization of uncertain parameters:
\begin{equation}
\label{eq:general_formulation}
    \min_{\boldsymbol{x}} \ \mathcal{R} \left[ f(\boldsymbol{x}, \boldsymbol{\mathrm{y}}) \right].
\end{equation}
For a given realization of uncertainty $\boldsymbol{\mathrm{y}}$, the total operating cost corresponds to the optimal value of a second-stage operational problem, in which the VPP adjusts its dispatch, reserve activation, and imbalance positions subject to financial commitments and physical feasibility:
\begin{equation}
\label{eq:second_stage_problem}
\begin{aligned}
    f(\boldsymbol{x}, \boldsymbol{\mathrm{y}}) = & \min_{\boldsymbol{z}} & & \mathcal{C}(\boldsymbol{x}, \boldsymbol{z}, \boldsymbol{\mathrm{y}}) \\
    & \text{s.t.} & & \text{Financial constraints} \\
    & & & \text{Physical constraints,}
\end{aligned}
\end{equation}
where $\mathcal{C}(\cdot)$ represents the total operating cost, including DER dispatch cost, imbalance penalties, and market revenues. The financial constraints enforce consistency between market commitments and link them to the VPP dispatch. Physical constraints ensure the feasibility of grid and DER operations. The financial and physical constraint sets are detailed in Section~\ref{sec:FinantialModel} and~\ref{sec:PhysicalModelPart}, respectively.

To solve \eqref{eq:general_formulation}, we approximate the distribution of $\boldsymbol{\mathrm{y}}$ by a finite set of $N^s$ scenarios $\mathcal{S}$ with realizations $\boldsymbol{y}_s$ and associated probabilities $\pi_s$. The two-stage problem is then reformulated as a deterministic problem over all scenarios. The nested $\min$-$\min$ structure is resolved by introducing scenario-specific second-stage variables $\boldsymbol{z}_s \in \mathbb{R}^{N^z}$ and replacing $f(\boldsymbol{x}, \boldsymbol{y}_s)$ with the corresponding explicit cost function and constraints.

When the VPP operator is assumed to be risk-neutral, uncertainty is handled by minimizing the expected value of the operating cost, i.e., $\mathcal{R}[\cdot] = \mathbb{E}[\cdot]$. This yields the following deterministic equivalent formulation:
\begin{equation}
\label{eq:risk_neutral_reformulation}
\begin{aligned}
    \min_{\boldsymbol{x}, \boldsymbol{z}_s} \quad & \sum_{s \in \mathcal{S}} \pi_s \mathcal{C}(\boldsymbol{x}, \boldsymbol{z}_s, \boldsymbol{y}_s) \\
    \text{s.t.} \quad & \text{Financial constraints } & \forall s \in \mathcal{S} \  \\
                      & \text{Physical constraints } & \forall s \in \mathcal{S}.
\end{aligned}
\end{equation}
Note that $\boldsymbol{x}$ is shared across all scenarios, satisfying the non-anticipativity condition.

\begin{figure}
    \centering
    \includegraphics[width=\linewidth]{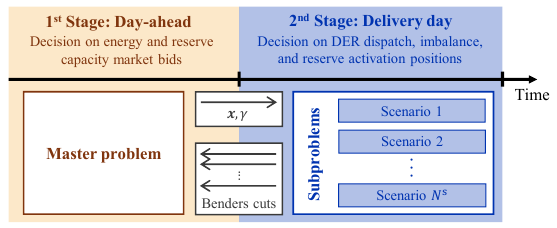}
    \caption{Visual representation of the decomposed problem, showing the master problem, scenario subproblems, and the exchange of information in Benders decomposition.}
    \label{fig:benders_visual}
\end{figure}

When the VPP operator is assumed to be risk-averse, uncertainty is handled by minimizing the Conditional Value-at-Risk (CVaR) of the operating cost, i.e., $\mathcal{R}[\cdot] = \mathrm{CVaR}_\alpha[\cdot]$, where $\alpha \in (0,1)$ is the considered confidence level. CVaR is defined as the expected value of the upper tail of the cost distribution beyond the $\alpha$-quantile. Therefore, minimizing CVaR shifts extreme cost realizations toward lower values, reducing exposure to adverse scenarios. The CVaR is defined as:
\begin{equation}
\label{eq:cvar_definition}
\mathrm{CVaR}_\alpha \left[ f(\boldsymbol{x}, \boldsymbol{\mathrm{y}}) \right] = \min_{\gamma \in \mathbb{R}} \left\{ \frac{1}{1 - \alpha} \mathbb{E} \left[ (\gamma - f(\boldsymbol{x}, \boldsymbol{\mathrm{y}}))_+ \right]- \gamma  \right\},
\end{equation}
where $\gamma$ is an auxiliary variable representing the $\alpha$-quantile of the cost distribution.
Under the scenario-based approximation, the CVaR objective is expressed in an equivalent linear form:
\begin{equation}
\label{eq:cvar_reformulation}
\begin{aligned}
    \min_{\boldsymbol{x}, \boldsymbol{z}_s, \gamma, y_s^{\mathrm{aux}}} \quad & \frac{1}{1 - \alpha} \sum_{s \in \mathcal{S}} \pi_s y_s^{\mathrm{aux}} - \gamma \\
    \text{s.t.} \quad & y_s^{\mathrm{aux}} \geq \gamma - \mathcal{C}(\boldsymbol{x}, \boldsymbol{z}_s, \boldsymbol{y}_s) & \forall s \in \mathcal{S} \ \\
                      & y_s^{\mathrm{aux}} \geq 0 & \forall s \in \mathcal{S} \ \\
                      & \text{Financial constraints } & \forall s \in \mathcal{S} \ \\
                      & \text{Physical constraints } & \forall s \in \mathcal{S} ,
\end{aligned}
\end{equation}
where $y_s^{\mathrm{aux}}$ is the non-negative auxiliary variable capturing the excess of each scenario cost above $\gamma$. 
Benders decomposition is adopted to exploit the two-stage structure of the problem, in which first-stage market bids are coupled across scenarios, while second-stage DER dispatch and network constraints are scenario-specific. This separation enables the decomposition of the large-scale stochastic program into a master problem and independent scenario subproblems, which can be solved in parallel. The master problem optimizes the first-stage variables $\boldsymbol{x}$ (or $\gamma$, $\boldsymbol{x}$ in the CVaR case), while scenario-specific subproblems solve for the second-stage variables $\boldsymbol{z}_s$ and compute scenario costs. At each iteration, Benders cuts (linear approximations of the recourse function) are added to refine the master problem’s feasible region until convergence is achieved. Compared to a monolithic formulation, this approach significantly reduces memory requirements and enables parallel computation across scenarios. Therefore, the model scales to detailed DER and grid representations with large scenario sets that would otherwise be computationally prohibitive. The structure of the decomposed problem is detailed in Figure~\ref{fig:benders_visual}.

\subsection{Financial constraints}
\label{sec:FinantialModel}
The VPP’s total net cost in each scenario $s \in \mathcal{S}$ is defined as the difference between all incurred costs and generated revenues. The total cost function $\mathcal{C}_{s}$ comprises revenues from energy and reserve markets, operations costs, network tariffs, and imbalance costs, and is expressed as:

\begin{equation}
\label{eq:TotalCost}
\begin{aligned}
\mathcal{C}_{s} = & - \left( \mathcal{R}^{\mathrm{DAM}}_{s} + \mathcal{R}^{\mathrm{RCM}}_{s} + \mathcal{R}^{\mathrm{RAM}}_{s} \right) \\
& + \mathcal{C}^{\mathrm{OPS}}_{s} + \mathcal{C}^{\mathrm{Tariff}}_{s} + \mathcal{C}^{\mathrm{IMB}}_{s},
\end{aligned}
\end{equation}
where $\mathcal{R}^{\mathrm{DAM}}_{s}$, $\mathcal{R}^{\mathrm{RCM}}_{s}$, and $\mathcal{R}^{\mathrm{RAM}}_{s}$ are the revenues on the day-ahead, reserve capacity, and reserve activation markets, respectively. Then, $\mathcal{C}^{\mathrm{OPS}}_{s}$, $\mathcal{C}^{\mathrm{Tariff}}_{s}$, and $\mathcal{C}^{\mathrm{IMB}}_{s}$ are the DER operations, tariff and imbalance costs.
To define the revenues, we divide the delivery horizon into a set $\mathcal{T}$ of time steps of duration $\Delta t$, which is the time resolution adopted for DER operations, day-ahead, reserve activation, and imbalance markets. 
Therefore, the day-ahead market revenue is:
\begin{equation}
\label{eq:RevenueDAM}
\mathcal{R}^{\mathrm{DAM}}_{s} = \sum_{t \in \mathcal{T}} \rho^{\mathrm{DAM}}_{t,s} \cdot P^{\mathrm{DAM}}_{t} \cdot \Delta t,
\end{equation}
where $\rho^{\mathrm{DAM}}_{t,s}$ is the DAM price at time $t$ and scenario $s$, and $P^{\mathrm{DAM}}_{t}$ is the DAM offered volume at time $t$.

Reserve capacity products have a specified duration $\Delta t ^{\mathrm{RCM}}$. We divide the delivery day into the set of time windows $\mathcal{T}^{\mathrm{RCM}}$ of duration $\Delta t ^{\mathrm{RCM}}$. The reserve capacity market revenue is:
\begin{equation}
\label{eq:RevenueRCM}
\mathcal{R}^{\mathrm{RCM}}_{s} = \sum_{\tau \in \mathcal{T}^{\mathrm{RCM}}} \rho^{\mathrm{RCM,up}}_{\tau,s} \cdot P^{\mathrm{RCM,up}}_{\tau} + \rho^{\mathrm{RCM,dn}}_{\tau,s} \cdot P^{\mathrm{RCM,dn}}_{\tau},
\end{equation}
where $\rho^{\mathrm{RCM,up}}_{\tau,s}$ and $\rho^{\mathrm{RCM,dn}}_{\tau,s}$ are the upward and downward RCM prices in time window $\tau$ and scenario $s$, and $P^{\mathrm{RCM,up}}_\tau$ and $P^{\mathrm{RCM,dn}}_\tau$ are the RCM offered volumes in $\tau$.

The reserve activation market revenues are:
\begin{equation}
\label{eq:RevenueRAM}
\begin{aligned}
\mathcal{R}^{\mathrm{RAM}}_{s} = & \sum_{t \in \mathcal{T}} \left( \rho^{\mathrm{RAM,up}}_{t,s} \cdot P^{\mathrm{RAM,up}}_{t,s} \right. \\
& \left. + \rho^{\mathrm{RAM,dn}}_{t,s} \cdot P^{\mathrm{RAM,dn}}_{t,s} \right) \cdot \Delta t,
\end{aligned}
\end{equation}
where $\rho^{\mathrm{RAM,up}}_{t,s}$ and $\rho^{\mathrm{RAM,dn}}_{t,s}$ are the upward and downward RAM prices at time $t$ and scenario $s$, and $P^{\mathrm{RAM,up}}_{t,s}$ and $P^{\mathrm{RAM,dn}}_{t,s}$ are the RAM offered volumes at time $t$ and scenario $s$. Reserve activations must not exceed the prequalified power \(\overline{P}^{\mathrm{RAM}}\):
\begin{align}
\label{eq:PrequalificationConstraint}
P^{\mathrm{RAM,up}}_{t,s} &\le \overline{P}^{\mathrm{RAM}}\\
P^{\mathrm{RAM,dn}}_{t,s} &\le \overline{P}^{\mathrm{RAM}}.
\end{align}
Moreover, it must be guaranteed that the reserved capacity is offered in the activation market; therefore, we enforce:
\begin{align}
\label{eq:RCMMappingUP}
P^{\mathrm{RAM,up}}_{t,s} & \ge P^{\mathrm{RCM,up}}_{\tau(t)} \\
\label{eq:RCMMappingDN}
P^{\mathrm{RAM,dn}}_{t,s} &\ge P^{\mathrm{RCM,dn}}_{\tau(t)},
\end{align}
where $\tau(t)$ maps each time element in $\mathcal{T}$ to the corresponding element in $\mathcal{T}^{\mathrm{RCM}}$.

The operational costs $\mathcal{C}^{\mathrm{OPS}}_{t,s}$ aggregate the expenses related to DER dispatch and customer compensation:
\begin{equation}
\label{eq:OperationalCosts}
\mathcal{C}^{\mathrm{OPS}}_{s} = \sum_{t \in \mathcal{T}} \mathcal{C}^{\mathrm{G}}_{t,s} + \mathcal{C}^{\mathrm{HP}}_{t,s} + \mathcal{C}^{\mathrm{EV}}_{t,s} + \mathcal{C}^{\mathrm{ST}}_{t,s},
\end{equation}
where $\mathcal{C}^{\mathrm{G}}_{t,s}$, $\mathcal{C}^{\mathrm{HP}}_{t,s}$, $\mathcal{C}^{\mathrm{EV}}_{t,s}$, and $\mathcal{C}^{\mathrm{ST}}_{t,s}$ are the operations costs of DGs, HPs, EVs, and BESS units, respectively. These costs are a function of the DERs' operations. For details, refer to~\cite{zapparoli2025powerreservecapacityvirtual}.

The network tariff costs are applied to the net power withdrawn from the grid at node $i \in \mathcal{N}$~\cite{ACER2023} as:
\begin{equation}
\label{eq:GridCost}
\mathcal{C}^{\mathrm{Tariff}}_{s} = \sum_{t \in \mathcal{T}} \sum_{i \in \mathcal{N}} P^{\mathrm{wit}}_{i,t,s} \cdot c^{\mathrm{tariff}}_{t},
\end{equation}
where $c^{\mathrm{tariff}}_{t}$ is the network tariff at time $t$, and $P^{\mathrm{wit}}_{i,t,s}$ is the power withdrawn at node $i$, at time $t$, in scenario $s$.

Finally, the imbalance costs reflect penalties or remunerations arising from deviations between contracted and actual energy deliveries, defined as:
\begin{equation}
\label{eq:ImbalanceCosts}
\mathcal{C}^{\mathrm{IMB}}_{s} = \sum_{t \in \mathcal{T}} C^{\mathrm{IMB,short}}_{t,s} - C^{\mathrm{IMB,long}}_{t,s},
\end{equation}
where $C^{\mathrm{IMB,short}}_{t,s}$ and $C^{\mathrm{IMB,long}}_{t,s}$ denote the costs and revenues associated with short and long imbalance positions, respectively. These costs are linear functions of the imbalance quantities, following a dual pricing mechanism:
\begin{align}
C^{\mathrm{IMB,short}}_{t,s} &= \phi^{\mathrm{short}}_{t,s} \cdot P^{\mathrm{IMB,short}}_{t,s}, \label{eq:IMB_short_simple} \\
C^{\mathrm{IMB,long}}_{t,s}  &= \phi^{\mathrm{long}}_{t,s}  \cdot P^{\mathrm{IMB,long}}_{t,s}, \label{eq:IMB_long_simple}
\end{align}
where \(\phi^{\mathrm{short}}_{t,s}\) and \(\phi^{\mathrm{long}}_{t,s}\) are the short and long imbalance prices. These coefficients are determined according to~\cite{Imbalance}.
Finally, we ensure the physical position of the VPP matches its financial positions in markets that require energy delivery:
\begin{equation}
\label{eq:CombinePositions}
\begin{split}
    P^{\text{VPP}}_{t,s} = P^{\text{DAM}}_t  + P^{\text{RAM,up}}_{t,s} - P^{\text{RAM,dn}}_{t,s} \  \\ -  P^{\text{IMB,short}}_{t,s} + P^{\text{IMB,long}}_{t,s},
\end{split}
\end{equation}
The imbalance position links the physical and financial positions and allows adjustments to be made to the first-stage decisions for the day-ahead and reserve capacity markets.

\subsection{Physical constraints}
\label{sec:PhysicalModelPart}
In this section, we describe the physical model, which consists of all the constraints defining the operations of the DERs and the network constraints.

\textit{Network constraints.} 
The network is modeled in all scenarios using the linear DistFlow equations~\cite{19266}. This formulation leverages the radial topology assumed for the VPP distribution network to provide a direct linearized version of the power flow equations. These constraints relate to the power injected by every DER to $P_{t,s}^{\text{PCC}}$ and model voltage and flow limits of the grid.
Voltage and flow limits are enforced for each bus and branch of the network. Voltage constraints are linear in DistFlow problems, while branch flow constraints, which would be quadratic, are piecewise linearized as in~\cite{zapparoli2025powerreservecapacityvirtual}. Lastly, nodal active and reactive power injections are linked to the power exchanges of the DERs through nodal balance constraints.

\textit{DER constraints}.
Single DERs are modeled using mathematical constraints that represent their operations. Different sets of constraints are applied to the four technology classes: DGs, HPs, EVs, and BESS units. For DGs, the real power output can be controlled from zero to the maximum power, which is computed as the product of the unit's nominal power and a capacity factor that models the primary resource availability. HPs are modeled to maintain comfortable indoor temperatures by controlling the heating power based on building properties (thermal resistance and capacitance) and ambient temperature. EVs are modeled based on charging events, considering vehicle-to-grid operations. BESS units and EVs are modeled with fixed charging and discharging efficiencies. The temperature for HPs and the state of charge for BESS units must be the same at the beginning and end of the observed period, ensuring the resources are available in the subsequent time steps. A minimum average charge rate is required for EVs. DGs and BESS units can exchange reactive power (limited by the inverter rating), whereas EVs and HPs cannot. For details on the DERs modeling, refer to~\cite{zapparoli2025powerreservecapacityvirtual}.
\begin{figure}
    \centering
    \includegraphics[width=\linewidth]{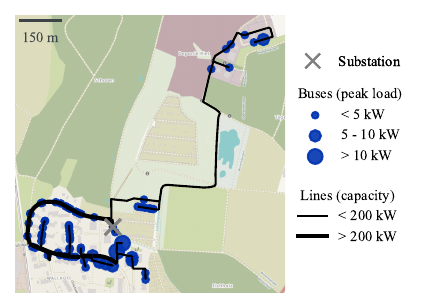}
    \caption{Map of the case study network (230-212\_1\_3), showing the 97 buses, lines, and substation locations.}
    \label{fig:network_map}
\end{figure}
\subsection{Uncertain parameters} 
The DGs' capacity factors, ambient temperature, EV behavior, non-dispatchable load, day-ahead reserve capacity, reserve activation, and imbalance market prices are uncertain at the decision time. Thus, they define the uncertainty set.

\section{Case study}
\subsection{Case study description}
\label{sec:cstudy_description}
The proposed framework is tested on a VPP based on a low-voltage network in northern Switzerland. The network topology is generated in~\cite{Oneto_2023} using open-source data, and~\cite{zapparoli2025futuredeploymentflexibilitydistributed} provides the installed DERs and their operations and flexibility for each node. The network is shown in Figure~\ref{fig:network_map}. It comprises 97 buses, with \(\SI{150}{kW}\) of peak non-dispatchable load, \(\SI{150}{kWp}\) of rooftop PV generation, \(\SI{85}{kW}\) of installed HPs, \(\SI{75}{kWh}\) of BESS, and 40 EV charging events during the considered time horizon. The EVs have an average battery capacity of \(\SI{70}{kWh}\) and an average maximum charging power of \(\SI{7}{kW}\).

The model uses publicly available Swiss day-ahead~\cite{entsoe_transparency}, secondary and tertiary reserve activation prices~\cite{swissgrid_tenders}. Since reserve capacity prices are not publicly accessible for Switzerland, this study uses data from the German market as a proxy~\cite{regelleistung2025}. Network and system tariffs \(c_t^{\text{tariff}}\) are set at \(\SI{206.5}{CHF/MWh}\)~\cite{ewz_2024_tariff}. For all historical data, 2024 values are used. We select March 13th as the simulation day. This day reflects typical spring or autumn conditions in which all DERs are active.

\begin{table}
\caption[Parameters for forecasting errors]{Uncertainty model parameters for the considered forecasting errors}
\label{tab:Forecasting errors}
\vspace{6pt}
\centering
\begin{tabular}{|c|c|c|c|c|} 
\hline
\textbf{Error} & \textbf{PDF} & \textbf{Standard deviation} & \textbf{Mean} & \textbf{Ref.} \\
\hline
$E^{\mathrm{L}}$ & Normal & 10.75\% & 0 &~\cite{Kong} \\  
$E^{\mathrm{G}}$ & Normal & 8.15\% & 0 &~\cite{zapparoli2025powerreservecapacityvirtual}\\  
$E^{\mathrm{T}}$ & Normal & 1.5\,K & 0 &~\cite{zapparoli2025powerreservecapacityvirtual}\\    
$E^{\mathrm{EV}}$ & Uniform & 5.77\% & 10\% &~\cite{zapparoli2025powerreservecapacityvirtual}\\ 
$E^{\mathrm{DAM}}$ & Normal & 4.28\,EUR/MWh & 0 &~\cite{Lago} \\  
$E^{\mathrm{RCM,aFRR}}$ & Normal & 3.30\,EUR/MW & 0 &~\cite{Cardo-Miota2023}\\  
$E^{\mathrm{RAM,aFRR,up}}$ & Normal & 32.08\,EUR/MWh & 0 &~\cite{Failing2025} \\ 
$E^{\mathrm{RAM,aFRR,dn}}$ & Normal & 21.25\,EUR/MWh & 0 &~\cite{Failing2025} \\ 
$E^{\mathrm{RAM,mFRR,up}}$ & Normal & 63.6\,EUR/MWh & 0 &~\cite{Failing2025}\\ 
$E^{\mathrm{RAM,mFRR,dn}}$ & Normal & 42.91\,EUR/MWh & 0 &~\cite{Failing2025}\\
\hline
\end{tabular}
\end{table}
\subsection{Uncertainty modeling}
\label{sec:cstudy_uncertainty_modeling}
Key non-financial input parameters subject to forecasting uncertainty include ambient temperature (which affects HP operation), solar irradiance (which influences photovoltaic generation), user behavior (which impacts EV usage), and baseline load profiles. Given the spatial proximity of the DERs and the short-term focus, we adopt the simplifying assumption that each type of forecasting error affects all corresponding DERs equally and consistently across all time steps. This allows us to model uncertainty in DER behavior using four lumped error parameters: $E^{\mathrm{G}}$ for solar generation, $E^{\mathrm{T}}$ for ambient temperature, $E^{\mathrm{EV}}$ for EV usage, and $E^{\mathrm{L}}$ for load profiles~\cite{zapparoli2025powerreservecapacityvirtual}.

For the financial input parameters, prices in the day-ahead and ancillary service markets, we assume that the associated forecasting errors are statistically independent from both the non-financial parameters and from each other, resulting in the lumped error parameters $E^{\mathrm{DAM}}$, $E^{\mathrm{RCM, aFRR}}$, $E^{\mathrm{RAM, aFRR}}$, and $E^{\mathrm{RAM, mFRR}}$.
All the forecasting errors are modeled as normally distributed random variables with zero mean, reflecting the assumption of unbiased forecasting errors~\cite{SRINIVASAN}. The error term for tertiary reserve activation $E^{\mathrm{RAM, mFRR}}$ is only used to model the uncertainty in imbalance costs. We derive all standard deviations used to parameterize the forecast error distributions from a comprehensive literature review. Table~\ref{tab:Forecasting errors} summarizes the resulting values. 

\subsection{Computational aspects}
\label{sec:cstudy_computational_aspects}
The flexibility assessment model is implemented in Python, using Pyomo as a parser and Gurobi as a solver. Benders decomposition is implemented through the open-source Python package \textit{mpi-sppy}. The model is executed on the ETH Zurich \textit{EULER} high-performance computing cluster. The execution is parallelized using MPI on a node with 48 cores and $\SI{48}{GB}$ of RAM. We generated $N^s=1000$ second-stage scenarios by using Latin hypercube sampling on the error distributions. We adopted a confidence level $\alpha = 0.9$ for the CVaR optimization. The risk-neutral problem with $\sim~$41.5~$\cdot10^6$ variables, and $\sim$~186.2~$\cdot10^6$ constraints takes \(\sim \SI{1}{h}~\SI{40}{min}\) to solve with Benders decomposition. The risk-averse problem is solved in a similar time. In contrast, the non-decomposed extensive-form formulation cannot be solved for $N^s=1000$ scenarios due to memory limitations, as it exceeds the available $\SI{4}{TB}$ of RAM on the computing cluster node.
\begin{figure}
    \centering
    \begin{tikzpicture}
        \node[anchor=south west, inner sep=0] (img) at (0,0)
            {\includegraphics[width=\linewidth]{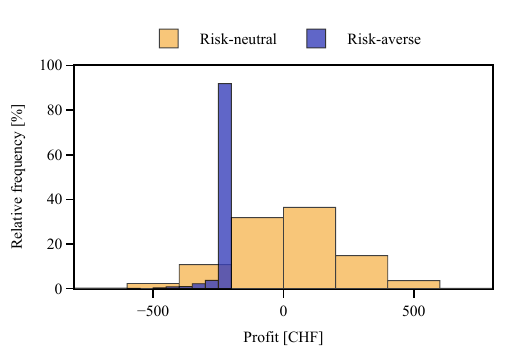}};

        \begin{scope}[x={(img.south east)}, y={(img.north west)}]

            \draw[gray, dashed, very thick] (0.431,0.205) -- (0.431,0.817);
            \node[gray, font=\scriptsize, anchor=east]
            at (0.431,0.75) {\SI{-242.3}{CHF}};

            \draw[gray, dashed, very thick] (0.566,0.205) -- (0.566,0.817);
            \node[gray, font=\scriptsize, anchor=west]
            at (0.566,0.75) {\SI{22.1}{CHF}};

        \end{scope}
    \end{tikzpicture}
    \caption{Distribution of VPP net profits under risk-neutral and risk-averse (CVaR) strategies. Dashed gray lines indicate the expected profit of each strategy.}
    \label{fig:profit_distribution}
\end{figure}
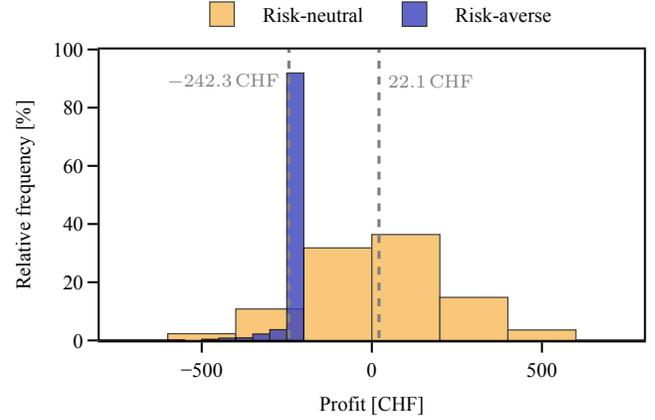

\section{Results and discussion}
\label{sec:results}

This section presents the case study results. First, Section~\ref{sec:results-risk} analyzes the effect of risk aversion on market participation and DER operation. Thereafter, Section~\ref{sec:results-tariff} evaluates the impact of dynamic network tariffs on the VPP operations and profitability.

\subsection{Impact of risk aversion}
\label{sec:results-risk}

Figure~\ref{fig:profit_distribution} shows the trade-off between profit and risk. The risk-neutral strategy (expected value optimization) yields an expected profit of \SI{22.1}{CHF} with a volatility of \SI{207.5}{CHF}, exposing the VPP to significant downside risk. In contrast, the risk-averse strategy (CVaR optimization) reduces volatility to \SI{31.0}{CHF} at the cost of an expected loss of \SI{242.3}{CHF}, accepting an expected profitability reduction of \SI{264.4}{CHF} in exchange for more stable financial outcomes.

\begin{figure*}
  \centering
  \subfloat[]{
    \includegraphics[width=0.48\textwidth]{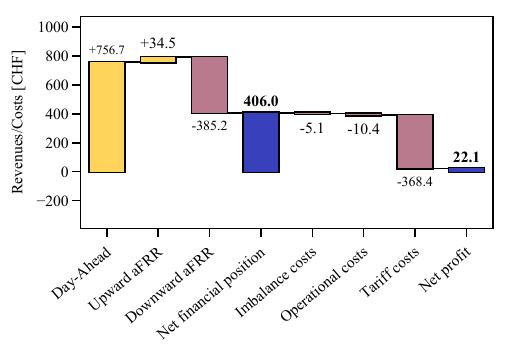}
    \label{fig:profit_maximizing}
  }
  \hfill
  \subfloat[]{
    \includegraphics[width=0.48\textwidth]{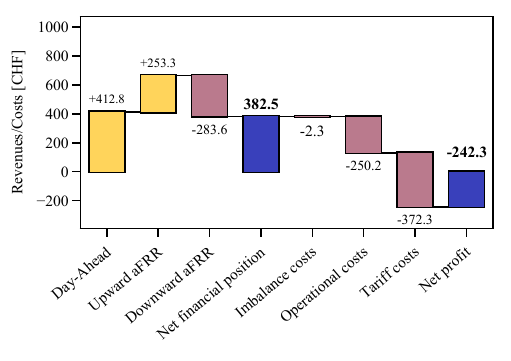}
    \label{fig:risk_averse}
  }
  \caption[Revenue and cost stream comparison]{Comparison of relevant expected revenue (positive) and cost (negative) streams for two bidding strategies. Panel (a) shows the risk-neutral strategy, and panel (b) illustrates the risk-averse strategy (CVaR optimization).}
  \label{fig:Cost_revenue_VPP}
\end{figure*}

Figure~\ref{fig:Cost_revenue_VPP} displays the expected daily profits of the two strategies, disaggregated into their expected revenue and cost components.
The risk-neutral strategy prioritizes short day-ahead positions and downward reserve activations to exploit expected price spreads. This aggressive positioning can yield high revenues when reserve activation and imbalance outcomes align with the market forecasts. However, it also results in a systematic mismatch between scheduled and physical quantities: the VPP schedules \SI{122.9}{kWh} in the day-ahead market, while only \SI{81.6}{kWh} are ultimately withdrawn at the PCC. The resulting deviations expose the VPP to imbalance penalties, explaining the high volatility of realized profits.
In contrast, the risk-averse strategy diversifies across markets, reducing exposure to volatile reserve prices. It achieves this by adopting both short and long reserve activation bids and by scheduling day-ahead positions that are more closely aligned with physical dispatch. 
Operationally, the strategy relies more heavily on internal flexibility across the DER portfolio. Most importantly, BESS throughput increases by \SI{146}{\percent} compared to the risk-neutral case. EV cycled energy also increases by \SI{7}{\percent}, while HP electricity consumption rises by \SI{2}{\percent}, as heat pumps operate at slightly higher average temperatures to create additional thermal flexibility. PV operation remains unchanged. This additional flexibility is used to hedge against imbalance penalties, shifting them from a mean of \SI{5.1}{CHF} with a volatility of \SI{16.1}{CHF} under the risk-neutral case to a reduced mean of \SI{2.3}{CHF} and volatility of \SI{9.13}{CHF} under the risk-averse case.

Finally, although the optimization model includes the reserve capacity market, the VPP does not place any bid. Consequently, the associated cash flow is zero. Therefore, the RCM is omitted from Figure~\ref{fig:Cost_revenue_VPP} for conciseness. This outcome reflects the restrictive four-hour booking duration and unfavorable price incentives, which make committing reserve capacity unattractive under operational uncertainty.

\subsection{Impact of dynamic network tariffs}
\label{sec:results-tariff}

\begin{figure*}
  \centering
  \subfloat[]{
    \includegraphics[width=0.48\textwidth]{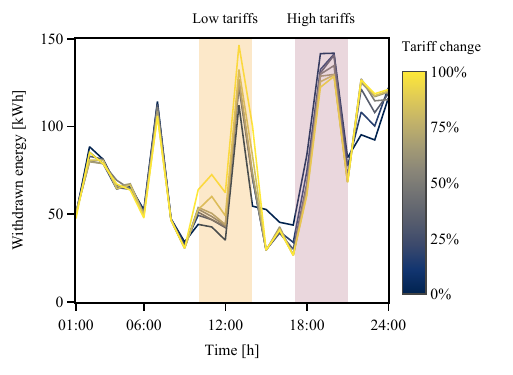}
    \label{fig:Dyn_Tariff_withdrawls}
  }
  \hfill
  \subfloat[]{
    \includegraphics[width=0.48\textwidth]{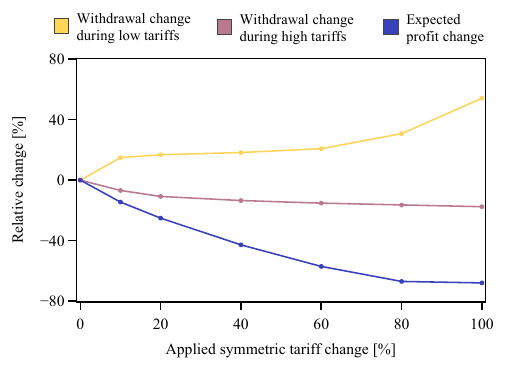}
    \label{fig:Dyn_Tariff_impact}
  }
  \caption[Dynamic tariff effects]{Impact of dynamic network tariffs on the VPP. 
    Panel~(a) illustrates the net energy withdrawn at the point of common coupling (PCC) under varying symmetric tariff change levels. 
    Blue–yellow colors correspond to the percentage of tariff variation applied, where shaded bands highlight low- and high-tariff periods (10~a.m.–2~p.m. and 5~p.m.–9~p.m., respectively). 
    Panel~(b) shows the relative change in expected profit and net withdrawals with respect to the baseline case without dynamic tariffs, under increasing tariff variation levels.}
  \label{fig:Dyn_Tariff_combined}
\end{figure*}

This analysis evaluates the risk-neutral VPP’s response to a symmetric dynamic tariff structure with reduced grid charges from 10~a.m. to 2~p.m. and increased charges from 5~p.m. to 9~p.m. These price signals aim to incentivize intraday load shifting by encouraging electricity consumption during midday and discouraging it during the evening peak~\cite{Federal_Office_of_Justice}. We conduct a sensitivity analysis to assess the operational and financial performance of the VPP. The magnitude of these tariff changes varies symmetrically from 10\% to 100\%. Results are reported in Figure~\ref{fig:Dyn_Tariff_combined}.

Figure~\ref{fig:Dyn_Tariff_withdrawls} shows that the VPP effectively responds to the dynamic tariffs: electricity withdrawals increase during low-tariff hours and decrease during high-tariff hours. Outside the defined windows, the VPP rebalances its operations to maintain nearly constant total daily withdrawals.
Figure~\ref{fig:Dyn_Tariff_impact} further shows that the demand response is nonlinear. The increase in consumption during low-tariff periods remains nearly constant for tariff reductions between 10\% and 60\%, but intensifies between 80\% and 100\%. Conversely, the reduction in demand during high-tariff periods saturates beyond a 20\% increase. Thus, moderate tariff variations are effective in shifting load, whereas stronger signals, above 20\%, induce limited additional flexibility while reducing VPP profitability by up to 65\%. These results highlight that modest tariff changes can deliver meaningful flexibility at limited cost, while excessive interventions erode economic performance with marginal operational gains.

\section{Conclusion}
\label{sec:conclusion}

This paper presents a two-stage stochastic optimization framework for the multi-market scheduling of DER-based Virtual Power Plants. The model integrates device-level constraints, network limitations, and large-scale uncertainty in market and operational variables, while embedding conditional value-at-risk to reflect risk preferences. Benders decomposition ensures tractability with detailed DER and grid models across extensive scenario sets, providing a scalable tool for co-optimizing market participation and response to dynamic network tariffs.

Results of a realistic case study show that both risk-neutral and risk-averse strategies exploit arbitrage opportunities across sequential markets. However, risk aversion reduces exposure to volatile reserve prices by aligning bids more closely with physical dispatch. This shift increases reliance on internal flexibility, particularly BESS operation, to hedge imbalance costs. Dynamic network tariffs further incentivize intraday load shifting. Moderate signals achieve meaningful flexibility at limited cost, while excessive signals erode profitability with marginal operational gains.

Limitations remain in the representation of uncertainty and market scope. The assumption of independent Gaussian forecast errors underestimates joint tail risks, which can be better captured with copula-based methods. Moreover, secondary reserve activation uncertainty can be explicitly modeled to enhance accuracy. Future work should explore alternative uncertainty-propagation techniques to further reduce the computational burden of large scenario sets while preserving accuracy. Finally, extending the framework to include the intraday energy market represents a key avenue for future work, given its growing role in system balancing and flexibility procurement.

\section*{Acknowledgements}
\label{Acknowledgements}
The research published in this paper was carried out with the support of the Swiss Federal Office of Energy (SFOE) as part of the SWEET consortium EDGE. The authors bear sole responsibility for the conclusions and results. The authors would like to thank Dr. Raphael Wu from Swissgrid AG for the support and insightful discussions.

\bibliographystyle{IEEEtran}
\bibliography{references}

\end{document}